\documentclass{emulateapj}
\begin{document}


\title{Dark matter cores in the Fornax and Sculptor dwarf galaxies: joining halo assembly and detailed star formation histories}


\author{N. C. Amorisco}
\affil{Dark Cosmology Centre, Niels Bohr Institute, University of Copenhagen, Juliane Maries Vej 30, 2100 Copenhagen, Denmark}

\author{J. Zavala}
\affil{Perimeter Institute for Theoretical Physics, 31 Caroline St. N., Waterloo, ON, N2L 2Y5, Canada\\
  Department of Physics and Astronomy, University of Waterloo, Waterloo, Ontario, N2L 3G1, Canada\\
  Dark Cosmology Centre, Niels Bohr Institute, University of Copenhagen, Juliane Maries Vej 30, 2100 Copenhagen, Denmark}

\author{T. J. L. de Boer}
\affil{Institute of Astronomy, University of Cambridge, Madingley Road CB3 0HA Cambridge, UK}




\begin{abstract}

We combine the detailed Star Formation Histories (SFHs) of the Fornax and Sculptor dwarf Spheroidals (dSphs) with the
mass assembly history (MAH) of their dark matter (DM) halo progenitors to estimate if the energy deposited by 
Supernova type II (SNeII) is sufficient to create a substantial DM core. Assuming the efficiency 
of energy injection of the SNeII into DM particles is $\epsilon_{\rm gc}=0.05$, we find that a {\it single}
early episode, $z\gtrsim z_{\rm infall}$, that combines the energy of all SNeII due to explode over 0.5 Gyr, is sufficient 
to create a core of several hundred parsecs in both Sculptor and Fornax. Therefore, our results suggest 
that it is energetically plausible to form cores in Cold Dark Matter (CDM) halos via
early episodic gas outflows triggered by SNeII. Furthermore, based on CDM merger rates and phase-space density 
considerations, we argue that the probability of a subsequent complete regeneration of the cusp is small for
a substantial fraction of dwarf-size haloes.

\end{abstract}





\section{Introduction}
Disagreements at small galactic scales are not a recent surprise for the $\Lambda$CDM paradigm, 
and the so called core-cusp problem is in fact one of the oldest. \citet{Flo94} and \citet{Moo94} first
brought the attention on the mismatch between the characteristic $\rho\sim r^{-1}$
density cusp observed in DM-only simulations \citep{Dub91, Nav96a} and the constant density cores 
inferred in dwarf disky galaxies. DM cores are now known to be ubiquitous in low surface brightness galaxies
\citep{Kuzio08} and nearby dwarf galaxies \citep{deBlok08}.
More recently, the problem has grown starker because of increasing, 
independent evidence that also several DM dominated dSphs 
may in fact host DM cores. The large orbits of the Globular Cluster system in Fornax 
\citep{Goe06, San06, Col12}, the survival of cold kinematical substructures in Ursa Minor and Sextans \citep{Kle03, Bat11}, and direct dynamical modelling of the stellar population 
of Fornax and  Sculptor \citep[see e.g.][]{Bat08, WaP11, Agn12, Amo13a}, all testify against the presence of 
divergent cusps in dSphs.

An emergent challenge is the so called `too big to fail' problem (TBTF). \citet{Boy11, Boy12}
(BK12) noticed that the observed 
central densities of the bright dSphs of the Milky Way (MW) are about a factor of a few smaller than those of 
the most massive satellites formed in DM-only cosmological simulations of MW-size halos. 

There are different means of easing these dwarf-scale controversies. Some of the proposed 
solutions require changes to the nature of DM as a particle. For example, non-zero thermal 
velocities imply a suppression of the DM power spectrum at subgalactic scales; however, in order 
to solve the core-cusp problem, a warm DM particle would be required to be unfeasibly warm \citep{Mac12}. 
DM collisionality, instead, might look like a more promising venue \citep{Vog12, Roc13, Zav13}, with 
self-interacting DM particles -- cross section of about $\sigma/m\gtrsim 0.6~{\rm cm}^2 {\rm g}^{-1}$ -- 
being able to better reproduce the central densities and core-sizes of dwarfs. 

However, it might not be necessary to abandon the realm of CDM to solve the core-cusp
problem. In fact, the first attempts at {\it transforming} a dwarf galaxy's cusp into a core
date back to \citet{Nav96b}. Through gravitational coupling, baryonic processes such as 
supernova (SN) momentum feedback could potentially be responsible for the cusp-core transformation.
By displacing the gas in essentially impulsive blow-outs { \citep{Pon12}}, bursts of SN explosions may be able to dynamically 
heat the central cusp, expanding the orbits of DM particles. Recent numerical experiments 
have observed such transformations in both cosmological hydrodynamical runs \citep{Mas08, Gov10, 
Zol12} and idealized controlled simulations \citep{Tey13}. 

Apart from more technical reasons connected with numerical resolution and/or with the different
implementations of the necessary sub-grid physics, some doubts still remain on the  
applicability of this scenario to the specific dSphs for which a core has been detected, e.g., Fornax and Sculptor.
For example,
BK12, \citet{Pen12} and \citet{GaK13} have warned that the energy necessary to reduce the central density of a
typical TBTF $\Lambda$CDM subhalo, with $V_{\rm max}\approx35$ kms$^{-1}$, to a level that is compatible 
with the bright MW satellites is in fact too large to be provided by stellar feedback. 

However, most of these calculations thus far have not taken into account the MAHs of such subhalos, whose bound 
mass has, on average, grown monotonically with redshift until accretion onto the MW. 
Additionally, in order to tailor such calculations to a specific dwarf, one should also take into account its particular 
SFH: feedback was most effective at specific moments of its assembly history, so that
different dwarfs will be impacted in diverse ways.

In this {\it Letter} we investigate in detail the specific cases of the Fornax and Sculptor dSphs, in order to ascertain whether
the observed cores in their DM density distributions can be the result of mechanical feedback from SN explosions.
We quantify the available energy by making use of their detailed SFHs \citep{deB12a, deB12b}. Hence, we explore 
the potential of such energy release on the likely progenitors at different redshifts. 

\section{Quantifying SN Rates through detailed Star formation Histories}

By using both deep multicolor photometry and large spectroscopic samples, 
\citet{deB12a,deB12b} have recently reconstructed the SFHs of the Fornax 
and Sculptor dSphs. They find these to be essentially different systems:
while Fornax experienced a complex SFH extending to less than only a few hundreds Myr ago, 
Sculptor had already completed its stellar build-up around 8 Gyr ago.
By integrating the star formation rate in time, we estimate that the total stellar mass of the Fornax 
dSph is $M_{*}=(3.12\pm0.35)\times 10^7 M_{\odot}$, while we obtain $M_{*}=(8.0\pm0.7)\times 10^6 M_{\odot}$ 
for the Sculptor dSph. 

\begin{figure}
\centering
\includegraphics[width=\columnwidth]{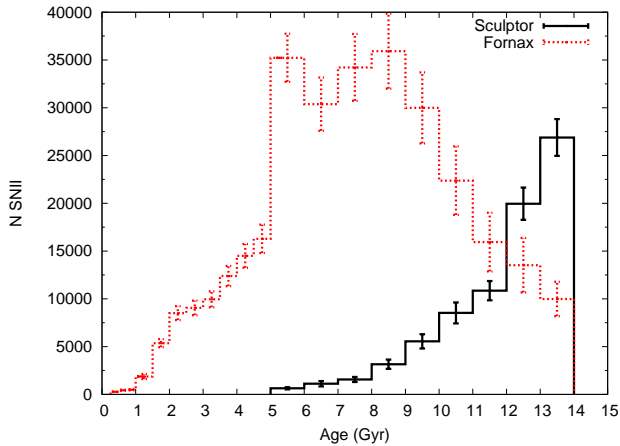}
\caption{The number of SNeII (per age bin) occurring in Sculptor~(black, solid) and Fornax~(red, dashed), according to the SFH as presented in \citet{deB12a,deB12b}. }
\label{SNII}
\end{figure}

We profit from the increased age resolution reached by \citet{deB12a,deB12b} 
and quantify the number of SN explosions at different moments in time. 
For simplicity, we only consider the explosion of SNeII, whose feedback 
dominates over that of SNeIa. Figure~1 displays the total number of SNeII explosions which 
occurred in each age bin of the SFH histogram. These are obtained by following the technique outlined in~\citet{Mat12}.
We isolate the fraction of the stars with mass in the interval $8<M/M_{\odot}<16$
that explode as SNeII -- rather than as type Ia -- and subsequently add the massive stars in the 
interval $16<M/M_{\odot}<40$. For a Kroupa IMF, this corresponds to about one SNeII
explosion for each 100 M$_{\odot}$. For Fornax, this implies that a total of 
$N_{SNII}\approx 3\times 10^5$ SNeII explosions occurred to the present day, while a total of $N_{SNII}\approx 8\times 10^4$ 
exploded in Sculptor over its history.

{ The DM halo only absorbs a small fraction of the total energy that is actually released in a burst of SNe. 
Part of the energy is first transferred to the gas, recent estimates place this efficiency between
5$\%$ \citep{Kir11} and a more generous 40$\%$ \citep{Gov10}. 
Then, rapid variations in the gravitational potential due to the gas displacement heat the 
DM particles. We adopt a phenomenological description of this process and parametrize the
total {\it gravitational coupling} with an efficiency $\epsilon_{\rm gc}$.} Also, despite great improvement
in the determination of SFHs, resolving in age an old burst of star formation is not yet possible. 
Henceforth, in order to translate the smooth rate of SNeII explosions $E'_{SN}$
(see lower panels in Fig.~2) into the actual strength of a particular burst $E_b$, 
we need to introduce a parameter for the `burstiness' of the SFH. We assume that the 
burst $E_b(t)=\int_{t}^{t+\tau} E'_{SN}(t') dt'$ collects the energy of SNe exploding over the timescale $\tau$.
In conclusion, the energy that is actually {\it injected} in the DM halo after one specific burst is
\begin{equation}
E_{\rm inj} = \epsilon_{\rm gc}\cdot E_b \approx \epsilon_{\rm gc}\cdot \tau\ E'_{SN} \ .
\label{effic1}
\end{equation} 
%

\section{Halo assembly histories and core-formation}

\citet{Pen12} presented a simple and efficient method to quantify the
energetics of the cusp-core transformation. Given an initial density profile, they 
used the virial theorem to estimate the minimum amount of energy that must be 
absorbed by the DM halo so that its particles can redistribute into a new equilibrium state:
\begin{equation}
E_{\rm inj} = (W_2-W_1)/2\ ,
\label{vth}
\end{equation} 
where $W$ is the total gravitational binding energy.
We assume that our initial states, $\rho_1(r)$, are cosmologically motivated cuspy NFW profiles \citep{Nav96a},
later to be replaced by a cored profile as a result of the energy injection:
\begin{equation}
\rho_2(r)={{\rho_{s, c}\ r_s^3}\over{(r+r_c)(r+r_s)^2}}\ ,
\label{cNFW}
\end{equation}
 where $r_c$ is the core radius and $\rho_{s, c}$ is such that the virial mass $M_{200}$ (the mass enclosed 
in a sphere with mean density 200 times the critical value) is conserved.

The mass assembly and structural evolution of $\Lambda$CDM halos has been followed closely by
countless numerical studies. { Although we are interested in subhaloes, we start by considering haloes in
isolation. }
\citet{Saw10,Saw11} and BK12 show that, for redshifts $z\lesssim10$, the median of the
bound mass is essentially exponential with redshift,
\begin{equation}
M_{200}(z)\approx M_{200}^{z=0}\ 10^{-z/z_0}\ .
\label{mgrowth}
\end{equation}
Additionally, comparing the same works we find that the characteristic time scale $z_0$ of such growth is 
approximately independent of $M_{200}^{z=0}$ in the interval we are interested in ($9\lesssim {\rm log}_{10}\left( M_{200}^{z=0}/M_{\odot}\right)\lesssim 10.5$), so that we can univocally fix it ($z_0^{-1}=0.15$). Mass growth is accompanied by structural evolution, which we can reproduce by evolving with redshift the
concentration-mass relation, for example as explicitly prescribed -- for isolated haloes -- 
by \citet{Gao08}. In conclusion, once the current mass of a $\Lambda$CDM halo $M_{200}^{z=0}$ has been chosen, we can reconstruct its assembly history and calculate, at any redshift, its detailed DM density profile, and then its binding energy $W_1$. 

Before addressing the specific cases of Fornax and Sculptor, it is useful to consider the
dimensionless ratio $E_{\rm inj}/W_1$. In terms of core-formation, interesting episodes of stellar feedback are
those in which the DM halo absorbs a fraction of its own potential energy: $0.01 \lesssim E_{\rm inj}/W_1 \lesssim 0.1$. 
When the injected energy is smaller, no astrophysically 
appreciable core is formed; if it is higher, the episode is potentially catastrophic in reshaping the halo, 
and the qualitative model given by eqn.~(2) likely breaks. 

The structural properties of $\Lambda$CDM halos are such that: 
\begin{equation}
W_1(z) \propto \rho_s^2\ r_s^5\sim M_{200}(z)^{1.65}\propto \left(M_{200}^{z=0}\right)^{1.65} \ ,
\label{wscal}
\end{equation}
i.e. there is a simple scaling between the potential energy $W$ and $M_{200}^{z=0}$, the only free parameter  
in the median MAH. As a consequence, since the SNeII rate $E'_{SN}$ is fixed by the SFH, 
the evolution of a given halo will be essentially driven by the global factor:
\begin{equation}
{{\epsilon_{\rm gc}\cdot \tau/{\rm Gyr}}\cdot {\left({M_{200}^{z=0}\over M_{\odot}}\right)^{-1.65}}} \equiv {{\epsilon_{\rm tot}}\cdot{\left({M_{200}^{z=0}\over M_{\odot}}\right)^{-1.65}} } \ .
\label{fudgef}
\end{equation}
In other words, given a burst of SN of energy $E_b(t)$, it is equivalent to have
a 4 times heavier halo or 10 times smaller total efficiency, both changes determine 
an analogous effect in the ratio $E_{\rm inj}/W_1$, corresponding to comparable results in terms of core-formation.

\subsection{Fornax and Scuptor}
{ Fornax and Sculptor did not evolve in isolation to present times. Infall
onto the MW halted their mass growth and stripped the outermost regions of their haloes.
This complicates the task of normalising their MAHs, parametrized by $M_{200}^{z=0}$. 
The mass at approximately the half-light radius $M^{z=0}(<R_h)$ represents the most robust 
dynamical constraint allowed by current kinematic data \citep{Wal09, Wol10, Amo12}, so
we identify compatible progenitors as those haloes that, at $z_{\rm inf}$, were 
massive enough to comply with the requirement 
\begin{equation}
M^{z=z_{\rm inf}}(<R_h) \gtrsim \alpha M^{z=0}(<R_h)\ .
\end{equation}

{ The correction $\alpha$ refers to the effects of core-formation on the central regions of the dwarf.
For cores with sizes of a few hundreds of parsecs, formed at $z\gtrsim 4$, we find $\alpha \lesssim 2$.
However, one may argue that cored subhaloes are susceptible to tides even near their centres \citep{Kaz13},
and that this effect should also be factored in $\alpha$. In practice, 
we find that a correction $\alpha\approx 2$ implies that both Fornax and Sculptor belong to the very high-mass end of
the distribution of TBTF subhaloes. While we cannot exclude this, we explicitly address this possibility in Sect.~3.2. 
For the moment, we prefer to adopt an educated estimate $\alpha\approx1.5$, which implies 
virial masses that are already a few times higher than allowed by kinematical constraints.}
As for the infall redshifts, a recent study estimates \citep{Roc12} that Fornax was accreted 4-9 Gyr ago, while
Sculptor 7-9 Gyr ago. Since higher infall redshifts imply higher virial masses, we conservatively use, for both dwarfs, 
the interval $1\lesssim z_{\rm inf}\lesssim2$,
which translates in the mass interval $9.4\leq {\rm log}_{10}(M_{200}^{z=0}/M_{\odot})\leq10$ for Fornax and 
$9.3\leq {\rm log}_{10}(M_{200}^{z=0}/M_{\odot})\leq9.9$ for Sculptor.

{ Figure~2 follows the MAHs of the Fornax and Sculptor dSphs (left and right panels, 
respectively) as a function of redshift and quantifies the effect of SNeII feedback by showing the `ratio of interest' $E_{\rm inj}/W_1$ 
of each possible SN burst $E_b(t)$, obtained from the SFHs.}
Shaded areas are obtained by assuming that the gravitational coupling efficiency is $\epsilon_{\rm gc}=0.05$ and that individual 
bursts collect the energy of the smooth SNe rate $E'_{SN}$ over a time $\tau=0.5$ Gyr. 
The width of the shaded areas in both panels reflects the uncertainty on the virial masses of Fornax and Sculptor.
{ The color-coding shows the core size that any of these bursts would individually form at a given redshift, starting from 
the corresponding $\Lambda$CDM cuspy progenitor, that is by ignoring any previous possible bursts.} Our results shows that at redshifts higher 
than 4 (6), a single burst of SNe with low coupling is indeed able of creating a core of at least 100 pc in Sculptor (Fornax). 
Lower virial masses in the allowed interval make the effect of feedback considerably more pronounced, with nominal core radii shooting above 1 kpc. 
The dashed red lines in both panels show the level of caution $E_{\rm inj}/W_1=0.2$, where eqn.~(2) likely fails. 
This might even be used to put upper bounds on the coupling efficiency $\epsilon_{gr}$ or on the strength of the burst itself (parameterised by $\tau$).

As prescribed by eqn.~(\ref{fudgef}), the vertical arrows in the right-hand panel of Fig.~2 indicate the magnitude of the effect of  
changes in the free parameters of the problem. In particular, given that subhaloes surviving to the present times 
might have been more concentrated than the average isolated halo with the same virial mass, we investigated the effect of
more concentrated progenitors. Higher concentrations make the cusp-core transformation
more challenging, although the effect is quite limited. A concentration that is two times higher than the
concentration-mass relation compiled by \citet{Gao08} (whose $95\%$ percentile is only 50$\%$ higher) 
results in a change in the core-sizes that is equivalent to a vertical shift of $\approx0.2$ dex in our Fig.~2, leaving our 
conclusions unchanged. 

\begin{figure*}
\centering
\includegraphics[width=\textwidth]{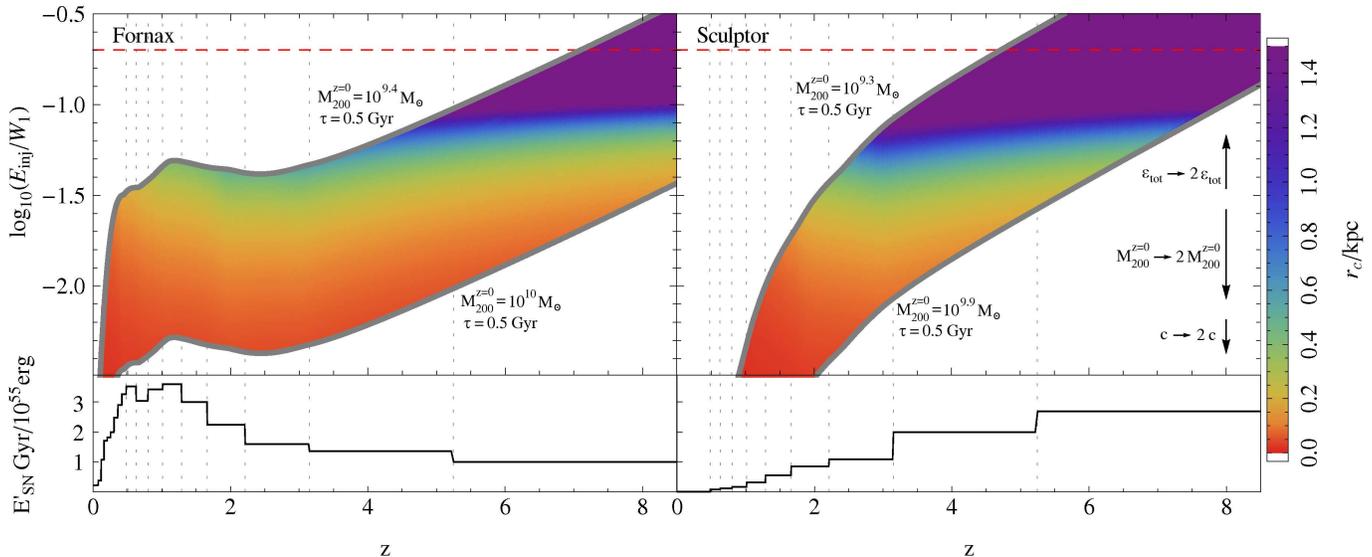}

\caption{The effect of SNeII momentum feedback obtained from the detailed SFHs of the Fornax (left panel) and 
Sculptor (right panel) dSphs. In both panels the shaded areas display the effect of individual bursts that collect 
{\it all} the energy of SNe exploding in a period of half a Gyr, for progenitors in the mass range $9.4\leq {\rm log}_{10}(M_{200}^{z=0}/M_{\odot})\leq10$ for Fornax, $9.3\leq {\rm log}_{10}(M_{200}^{z=0}/M_{\odot})\leq9.9$ for Sculptor.
The gravitational coupling of the bursts with the DM halo occurs with an efficiency of $\epsilon_{\rm gc}=0.05$. The lower panels display 
the history of SN rates according to the SFHs of Fig.~1. The arrows on the right panel indicate the magnitude of the vertical shifts that
occur if certain parameters are varied as given in the legends.}
\end{figure*}

Since eqn.~(4) ignores the effect of tidal stripping after infall, Fig.~2 underestimates the effect of 
any burst at $z\leq z_{\rm inf}$. This is especially relevant for Fornax, that has 
an important stellar population of intermediate age. As the arrows in the right-panel show, at $z\approx 1$, for a reduction of a factor of 4 of the 
bound mass, a single burst accumulating the energy of all SNe exploding over a 0.5~Gyr period
with a low coupling could be capable of forming a core of about 1 kpc.


\begin{figure}
\centering
\includegraphics[width=\columnwidth]{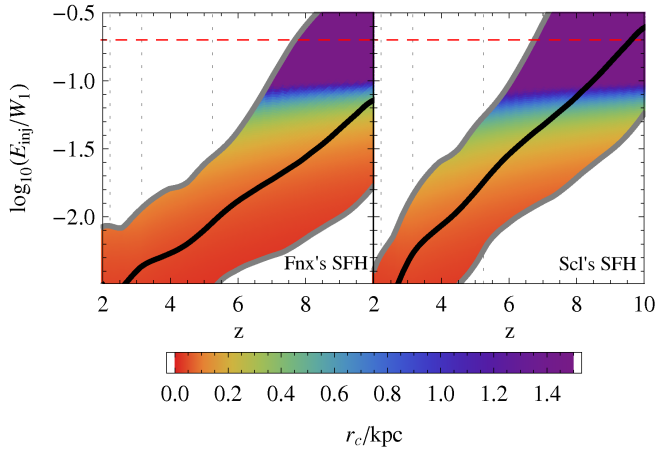}
\caption{The effect of the stellar feedback obtained from the SN rates of Fornax and Sculptor (left and right panels, respectively)
during the MAH of a typical `too big to fail' DM subhalo ($68\%$ confidence region of the MAHs of $V_{\rm infall}>30$ kms$^{-1}$ haloes, 
see BK12). Injected energies are calculated assuming a burst with $\epsilon_{\rm gc}=0.05$ and $\tau=0.5$~Gyr.}
\end{figure}

\subsection{The `too big to fail' problem}

{ A typical $\Lambda$CDM halo with $V_{\rm max}\approx35$ kms$^{-1}$ has a virial mass of $M_{200}^{z=0}\approx10^{10.2}M_{\odot}$. 
This is more similar to the mass-range suggested for Fornax and Sculptor by abundance-matching \citep[see][]{Pen12}, { while still lower than what we would obtain by accounting for the potential effects of tides on a shallow subhalo -- up to a factor of 8 in mass within the central kpc \citep{Kaz13}.}
We investigate the effect of the SFHs of Fornax and Sculptor on TBTF subhaloes by directly using the 
MAHs of massive subhaloes with $V_{\rm infall}>30$ kms$^{-1}$, recorded in Fig.~4 of
BK12.} For the $68\%$ confidence region in the mass assembly distribution, our Fig.~3 displays the 
evolution of the ratio of interest $E_{\rm inj}/W_1$ for the case of bursts with $\tau=0.5$ Gyr,
and a gravitational coupling efficiency $\epsilon_{\rm gc}=0.05$.
Despite the larger binding energies, the SFHs of both Fornax and Sculptor (left and right panels, respectively) 
prescribe enough SN explosions in their oldest couple of bins --
$t\gtrsim 12$ Gyr or $z\gtrsim3.5$ -- that a suitably early burst would be
able to enforce the formation of a core of several hundreds of parsecs. 
In order to lower the central density of a $V_{\rm max}\approx35$ kms$^{-1}$ $\Lambda$CDM halo 
to the levels of the MW bright dSphs (for example $V_{circ}\approx 15$ kms$^{-1}$ at $r=600$ pc), 
a core of about $r_c\approx1$ kpc is needed,
so that a {\it single} burst at redshifts $z_b\gtrsim10$ is necessary for Fornax, while $z_b\gtrsim7$ 
is likely enough for Sculptor.

\section{Discussion and Conclusions}

The detailed SFHs for the Fornax and Sculptor dSphs \citep{deB12a,deB12b} imply that
a significant number of SNeII exploded within their DM progenitors long before infall
onto the MW. By coupling these SNe rates with the MAHs of the DM progenitors, as given
by recent CDM $N-$body simulations, we find that it is energetically plausible to create cores of several hundred 
parsecs in both dwarfs. Sufficient conditions for this to happen are (i) an energy deposition from the SNe
into the DM particles with a coupling efficiency of $\epsilon_{\rm gc}=0.05$, and (ii) a {\it single} burst occurring before
infall that instantaneously collects the energy of all SNe due to explode over 
a period of about 0.5 Gyr. A higher coupling efficiency, a more violent burst
and/or repeated events would create even larger cores.

The earliest time when such a burst could have occurred is limited by photo-heating from the UV background after reionization, 
which prevents the condensation of gas in the early stages of the progenitor. The questions of how ``bursty'' the SFH can be, and how ``early'' the first effective burst can occur, remain open. Current implementations of galaxy formation simulations have not reached a consensus on the treatment of gas outflows driven by feedback, with different treatments seemingly producing realistic galaxy populations \citep[e.g.][]{Vog13,Mun13,Sta13,Mar13}. 

After the creation of such cores, the growth of the DM progenitors 
poses a threat to their survival, but complete reinforcement of the cusp { for the entire population of dwarfs} seems unlikely. 
Slow accretion and minor mergers
should deposit a substantial amount of cold material into the central regions, 
with low specific angular momentum, which is not what is observed
in the {\it inside-out} slow growth of CDM haloes. As for major mergers, because of 
phase-space density conservation, a dominant cusp can be reinforced if the
merging system is itself cusped \citep{Deh05} and the mass ratio
is sufficiently high. Merger rates are mass dependent in CDM, so that only about 50$\%$ of dwarf-size haloes 
like Fornax and Sculptor experience a merger with mass ratio $\gtrsim$1/3 between $z_{\rm inf}\approx1\leq z\leq4$ \citep{Fa2010}.

Our main finding is that the energy requirements for the cusp-core transformation in dSphs 
do not seem extremely demanding when compared to their detailed SFHs, if these were indeed bursty
{ and if they have not been significantly affected by tides.} 
Our results provide an additional motivation to look for observational signatures that can clearly distinguish a bursty from a quiescent SFH at time-scales $\lesssim0.1$~Gyr (observational evidence seemingly supporting bursty star formation has been presented e.g.~\citealt{vW11, Mas13}). 
For instance, for those dwarfs in which kinematically cold clumps are present, these can be used to put lower bounds on the age of any useful SN burst, which would otherwise pose a serious threat to the survival of such delicate substructures.

\acknowledgments
It is a pleasure to thank Niayesh Afshordi, Michael Boylan-Kolchin, James Bullock, Shea Garrison-Kimmel, Fabio Governato, 
Jorge Pe{\~n}arrubia, Matthew Walker for constructive discussions and the anonymous referee for helping us to improve the manuscript.
The Dark Cosmology Centre is funded by the DNRF.
JZ is supported by the University of Waterloo and the Perimeter Institute for Theoretical Physics. Research at Perimeter Institute is supported by the Government of Canada through Industry Canada and by the Province of Ontario through the Ministry of Research and Innovation.

\end{document}